\documentclass{amsart}
\usepackage{amssymb}
\def\preprint{}       
\def\finished{}
\def\archive {}
\newcommand{\cqd}{\hfill $\rule{2mm}{2mm}$}
\def\anel#1{$\langle #1,+,\cdot\rangle$}   
\newtheorem{teorema}{Theorem}
\newtheorem{define}{Definition}        
\newtheorem{lemma}{Lemma}
\long\def\abstract{ We will pursue a way of building up an algebraic structure that involves, in a mathematical abstract way, the well known Grassmann variables. The problem arises when we tried to understand the grassmannian polynomial expansion on the scope of ring theory. The formalization of this kind of variables and its properties will help us to have a better idea of some algebraic structures and the way they are implemented in  models concerning theoretical physics . \\
\begin{description}
	\item[{\bf Key-words}] Algebraic structures, Grassmann Numbers,  Non-commutative Algebras.	
\end{description}
	 
}
\input{epsf}
\begin{document}
\vskip 10mm
{\hfill \archive   \vskip -2pt \hfill\preprint }
\vskip 10mm
\centerline{\huge\bf An Attempt of Construction}
\vskip 4mm
\centerline{\huge\bf for the Grassmann Numbers.}
\vskip 7mm
\begin{center} \vskip 7mm
Ricardo Bent\'{\i}n{\begingroup\def\thefootnote{$\aleph$}
                                \footnote{e-mail: rbentin@uesc.br}
                                \addtocounter{footnote}{-1} \endgroup}\\                           
 Departamento  Ciências Exatas e Tecnológicas - DCET\\
 Universidade Estadual de Santa Cruz\\
 Rodovía Ilhéus-Itabuna km 16 s/n, CEP 45662-000, Ilhéus-BA, Brazil.      
\vfill {\bf ABSTRACT} \vskip 3mm \end{center}    \abstract

\vfill \noindent \\[5pt] \finished \vspace*{3mm}
\thispagestyle{empty} \newpage
\pagestyle{plain}
\newpage
\setcounter{page}{1}
\section{\sc Introduction.}
Let´s consider this:
\begin{define}[Grassmann Variable]
	\label{grass}
	Let $\mathbb{G}$ be a non-empty subset and let $\cdot$ and $+$ be two binary operations on this set such as:
	$$\mathbb{G}=\{\theta_i /\ \ \  \theta_i\cdot\theta_j+\theta_i\cdot\theta_j=0\},$$
	where $0$ is the additive identity of the complex ring. We define the set $\mathbb{G}$ as being the set
	of the Grassmann variables.
\end{define}
With this definition we start a study that will try to formalize a tool used in mathematical-physics \cite{liv}.\\
The idea of {\it variable} is in essence a way to represent our misunderstanding concerning some unknown fact and is this that we want to emphasis with the word {\it variable} in our starting definition.\\
Without going onto philosophical aspects, we can cite as an example the case of the polynomials defined over a field $K$ 
when they are written in its {\it usual} form in terms of the {\it variable} $x$,
$$p(x)=a_0+a_1x+\cdots+a_nx^n,$$
the set of these polynomials (together with the binary operations of sum and product defined for polynomials) describe an
structure called Euclidean Domain \cite{alg} and the notation used is $p(x)\in K[x]$. But this notation hides an important fact: 
we are actually working with an infinite sequence of elements that belongs to the algebraic structure: the field $K$.\\
Another example can be given by the equation
$$7x-3=1$$
when it is defined over the integer the $x$ {\it variable} is lacking any significance, meaningless, despite 
the other elements ($7$, $3$ and $1$) are well defined in $\mathbb{Z}$. Historically this is related with how we can find the roots
of a polynomial and in some higher level with the beauty (without a doubt) of the Galois Theory that relates the automorphisms over
a field extension.\\
These examples show us how some time something that is apparently trivial can intrinsically have fundamental concepts
such as the construction of rational numbers or the idea of the real numbers field.\\
But the Definition \ref{grass} is not at all completely consistent, since it let us some questions without answer or in some sense
ambiguity answers. For example, working with the set $\mathbb{G}$, will it be an extension of some another field (such as the field  $\mathbb{R}$ field or the $\mathbb{C}$ field)? Or may this kind of variables can be expressed in terms of known numerical sets? And 
what about the binary operations over the set $\mathbb{G}$, how can we define them?\\
In the next sections we will try to study if there are some implications on the definition concerning the Grassmann variable and we 
will try to make some constructions on them trying to use the fact that the additive identity of the complex field is present on 
Definition \ref{grass}.\\
The following section will contain a detailed exposition of the problem from the point of view of the the mathematical rigor 
and under the scope of the theoretical physics, when this kind of variables are used in a polynomial expansion. 
\section{The Problem.}
We are going to use the definition of Grassmann variable as it was given in the first section.\\
Works in theoretical physics consider that one fundamental particle in nature obey just one of the two possible statistics \cite{fis}:
\begin{itemize}
	\item Quantum Fermi-Dirac statistics.
	\item Quantum Bose-Einstein statistics.
\end{itemize}
This means that, in a mathematical language, we have two equivalence classes over the set of the fundamental particles: 
the set of {\it fermions} and the set of {\it bosons}.
It can be summed up as it is shown in Table \ref{fb:Class}.\\
\begin{table*}[h]
  \label{fb}
	\centering
		\begin{tabular}{|l|cc|}
			\hline
			Quantum statistics&Spin&Example of particles\\
			\hline\hline
			Fermi-Dirac&half-integer&leptons, quarks\\
			Bose-Einstein&integer&fotons, gluons\\
			\hline
		\end{tabular}
		\vskip 0.3mm
		\caption{Elementary particle classification according the statistical mechanics.}
	\label{fb:Class}
\end{table*}
To obtain a mathematical description of the quantum fermionic fields it is been used the Clifford Algebra (a ''slightly'' variant of Grassmann Algebra) \cite{fis,sup}. This statistical classification let us to arrive to the very {\it important concept of supersymmetry} in which the number of fermions is equal to the number of bosons. But when this concept is transcribed to the mathematical language, is mandatory the use of a new kind of variables to obtain a consistent supersymmetric space or say it in a short way: the superspace.\\
This superspace makes use of the definition of Grassmann variable and it is associated to the theoretical physcics in a bijective way:
one fermionic coordinate is represented by one Grassmann variable. This kind of construction is well known in mathematical-physics theories that involves supersymmetry \cite{sup} and since it is easy to see from the Definition \ref{grass} implies that $\theta^2=0$,
we can build up polynomials of the form:
\begin{equation}
\label{pol}
  p(\theta)=a_0+a_1\theta,
\end{equation}
where $a_0\in\mathbb{C}$ and $a_1,\theta\in\mathbb{G}$.\\
But a carefully observation allows us to see that there is a small inconsistency (from the mathematical point of view). To take account
of that, let´s remember how a polynomial is defined according the fundamentals of algebra:
\begin{define}[Polynomial] Let \anel{A} be a ring and $a_i\in A$ where  $i=0,\cdots,n$ and $n\in\mathbb{N}$. We define the finite sequence $\{a_0,\cdots,a_n\}$ of elements that belongs $A$ as being a polynomial of degree $n$ over the ring $A$.
\end{define} 
But in the polynomial of equation (\ref{pol}) we have that $a_0$ is over the ring of the complex numbers but $a_1$ does not belong 
to that ring at least. $a_1$ is on the set $\mathbb{G}$ and we does not have an idea of what is the algebraic structure that $\mathbb{G}$ belongs to! In order to have $\mathbb{G}$ as a ring structure we must have two binary operations. But our starting definition let us to think in a binary relation of the form $\mathbb{G}\times\mathbb{G}\mapsto\mathbb{C}$ that also restricts our idea of binary operation! In other words, we are in trouble when consider our Definition \ref{pol} under a mathematical lens.
\section{\sc Complex numbers and $2\times 2$ matrices.}
In this section we are going to describe, in a brief way, the relationship that exists between the complex numbers and the second order matrices. 
\begin{define}[Complex Number] Let $a$ and $b$ two elements that belong to the real numbers field $\mathbb{R}$, we define the element $z$ of the set $\mathbb{C}$ as $z=a+ib$, where is imposed the restriction $i^2=-1$. This set will be defined as the set of the complex numbers.
\end{define} 
\begin{define}[Operation $+$] Let $z_1=a_1+ib_1$ and $z_2=a_2+ib_2 \in \mathbb{C}$ two elements that belong to the set of complex numbers, we define the binary operation that we will call the addition  $(+)$ over $\mathbb{C}$ such as $z_1+z_2=(a_1+a_2)+i(b_1+b_2)$.
\end{define} 
\begin{define}[Operation $\cdot$] Let $z_1=a_1+ib_1$ and $z_2=a_2+ib_2 \in \mathbb{C}$ two elements that belong to the set of complex numbers, we define the binary operation that we will call the multiplication  $(\cdot)$ over $\mathbb{C}$ such as  $z_1\cdot z_2=(a_1a_2-b_1b_2)+i(a_1b_2+a_2b_1)$.
\end{define}
From here to now we are going to use the notation in which the product is implicit, i.e., $z_1\cdot z_2\equiv z_1z_2$. The folowing theorem defines one algebraic structure for the set $\mathbb{C}$ together with both former binary operations defined. 
\begin{teorema}[The Field $\mathbb{C}$] The set of the complex numbers together with the operations of addition and product describe an
algebraic structure of field.
\end{teorema}
\begin{description}
\item[Proof] Let $z_1=a_1+ib_1, z_2=a_2+ib_2, z_3=a_3+ib_3 \in \mathbb{C}$. Since we have the definition of $(+)$ it is easy to see that $\mathbb{C}$ form an Abelian group, and also that $z_1(z_2+z_3)=(a_1+ib_1)(a_2+ib_2)+(a_1+ib_1)(a_3+ib_3)=z_1z_2+z_1z_3$. then we just have to prove that for the operation $\cdot$, $\mathbb{C}$ also describes an Abelian group. Form the definition of the product we have that $z_1(z_2z_3)=a_1a_2a_3-a_1b_2b_3-b_1a_2b_3-b_1b_2a_3+i(a_1a_2b_3+a_1b_2a_3+b_1a_2a_3-b_1b_2b_3)=(z_1z_2)z_3$, that makes $\mathbb{C}$ a semigroup under the multiplication. But $1+i0 \in\mathbb{C}$ has the role of the identity for every element in $\mathbb{C}$, then we have that $\mathbb{C}$ is also a multiplicative monoid. In order to get a group structure, we have to find an inverse element for every $z=a+ib \in\mathbb{C},\ \ a\neq 0\neq b$, but $z^{-1}=(a-ib)/{a^2+b^2}$ implies that $zz^{-1}=z^{-1}z=1+i0$. Finally form the definition of the product it is trivial to see that $\mathbb{C}$ describes an Abelian group.\cqd
\end{description}
Now we will try to find an answer to the doubts concerning the equation \ref{pol}. For that we will use the result of the following theorem.
\begin{teorema}[Second Order Matrices] The subset $\mathcal{M}$ of the second order matrices that have the form:
$$
\begin{bmatrix}
a&-b\\
b&a
\end{bmatrix},
$$
where $a$ and $b$ represent any real numbers describe a field that is isomorphic to the field of the complex numbers.
\end{teorema}
\begin{description}
\item[Proof] These matrices describe a field. Since, under the matrix addition they form an Abelian group but also under the matrix product they describe an abelian group too. We only have to prove that the isomorphism exists.
Let the application $\varphi :\mathbb{C}\mapsto \mathcal{M}$ defined as:
$$
\varphi(z)=\varphi(a+ib)=\begin{bmatrix} a&-b\\ b&a \end{bmatrix},
$$
this aplication is a bijection an also preserve the operations of addition and product in $\mathbb{C}$ with the usual operations of addition and product of matrices in $\mathcal{M}$ since:
\begin{eqnarray*}
\varphi(z_1+z_2)&=&\varphi(a_1+a_2+ib_1+b_2)
=\begin{bmatrix}a_1+a_2&-b_1-b_2\\b_1+b_2&a_1+a_2\end{bmatrix}\\
&=&\begin{bmatrix}a_1&-b_1\\b_1&a_1\end{bmatrix}
+\begin{bmatrix}a_2&-b_2\\b_2&a_2 \end{bmatrix}
=\varphi(z_1)+\varphi(z_2),
\end{eqnarray*}
\begin{eqnarray*}
\varphi(z_1z_2)&=&\varphi(a_1a_2-b_1b_2+ia_1b_2+ia_2b_1)
=\begin{bmatrix}a_1a_2-b_1b_2&-a_1b_2-a_2b_1\\a_1b_2+a_2b_1&a_1a_2-b_1b_2\end{bmatrix}\\
&=&\begin{bmatrix}a_1&-b_1\\b_1&a_1\end{bmatrix}\begin{bmatrix}a_2&-b_2\\b_2&a_2 \end{bmatrix}
=\varphi(z_1)\varphi(z_2),
\end{eqnarray*}
then we can conclude that $\varphi$ is the desired isomorphism.\cqd
\end{description} 
Whit these results we are ready to go to the next section.
\section{\sc Grassmann Variables and $2\times 2$ Matrices.}
The main idea of this section is to search for some second order matrix representation that obeys the Definition  \ref{grass} with the condition (that belongs to the equation (\ref{pol})) that the operation of two Grassmannian variables led us to a complex number, that according the previous section is also a second order matrix. Let´s remember that our principal motivation is the fact that in general the product of two matrices is not commutative and if we find an appropriate matrix representation, the binaries operations over the set $\mathbb{G}$ will be automatically defined as the operations of addition and multiplication of second order matrices. The following result will help us in our analyze.  
\begin{lemma} Let $a,b\in\mathbb{R}$ both different from zero, the $2\times 2$ matrix that have the form
$
\begin{bmatrix}
ab&b^2\\
-a^2&-ab
\end{bmatrix}
$ fulfills  the Definition \ref{grass} but is unique unless real multiples of it.
\end{lemma}
\begin{description}
\item[Proof]  It is easy to see that the given matrix fulfills the definition of the Grassmann variables algebra, then let 
$$
\begin{bmatrix}
cd&d^2\\
-c^2&-cd
\end{bmatrix}
$$ 
another matrix that obeys the same Definition \ref{grass}, in this case these two matrices must also obey the same definition, but this implies the condition $(ad-bc)^2=0$, i. e., $ad=bc\doteq w\in\mathbb{R}$ that allows us to write $d=w/a$ e $c=w/b$, then the second matrix will be written as
$$\frac{w^2}{(ab)^2}
\begin{bmatrix}
ab&b^2\\
-a^2&-ab
\end{bmatrix}.
$$\cqd
\end{description}
This Lemma lead us to conclude that is not possible to construct a linearly independent set of second order matrices that respects the definition of Grassmann variables and also when multiplied the result is a {\it non-null} complex number associated to another second order matrix.
\section{\sc Grassmann Variables and the Complex Numbers.}
Since it is impossible to build a set of $2\times 2$ matrices that can be represented as the Grassmann variables of the Definition \ref{grass}, let´s try another possibility, now working with the idea of binary operations among the Grassmann variables. For this motive, we will give the following definitions,
\begin{define}[Odd Complex Function] Let $F(z):\mathbb{C}\mapsto\mathbb{C}$ a complex number application over the field $\mathbb{C}$, we define this function as an odd complex function if, and only if $F(z)=-F(-z)$ for all $z\in\mathbb{C}$.
\end{define}
\begin{define}[Binary Operation $*$]
\label{prod} Let $\theta_1=z_1$ and $\theta_2=z_2$ such us $z_1,z_2\in\mathbb{C}$ two elements of the complex numbers field, we define the Grassmann multiplication $(*)$ over $\mathbb{C}$ such us $\theta_1*\theta_2=F(z_1-z_2)$ where $F(z)$ is and odd complex function.
\end{define}
It is important to point out about this definition that the use of the $\theta$ variables only represent a label, since they are in fact elements of the complex number set, but now with the operation $*$ defined does not form a group, but a grupoid.\\
With this last Definition \ref{prod} of the binary operation $*$ we can shown that the Definition \ref{grass} is also valid.\\
But here we have to make some observations about it since the polynomial (\ref{pol}) has elements of the Grassmann type and also of the complex number type. Then it is necessary to define which binary operations can be used to relate $zR\theta$. In this case we have two choices:
$z\cdot\theta$ or $z*\theta$, taking in account physic´s considerations the suitable choose between $z$ and $\theta$ will be $(\cdot)$.\\
The use of an odd function when the Definition \ref{prod} was given have some interesting consequences on the scope of the mathematical-physics, since the non-commutative geometry \cite{sw} makes use of
$$[\theta_i,\theta_j]=i\Omega_{ij}$$
Where $[\theta_i,\theta_j]\equiv \theta_i*\theta_j-\theta_j*\theta_i$ represents the commutator, that when mixed with the definition of the $*$ product and the fact that $F(z)$ is odd lead us to the identification
$$\Omega_{ij}=-i2F(z_i-z_j)$$ that written in the form of the product $\star$, is known as the Moyal product \cite{sw} for two functions:
$$
	f(x)\star g(y)=e^{F(z_i-z_j)\partial_{x_i}\partial_{y_j}}f(x)g(y)\biggm|_{x=y},
$$
or using the Definition \ref{prod}
\begin{equation}
	f(x)\star g(y)=e^{\theta_i\theta_j\partial_{x_i}\partial_{y_j}}f(x)g(y)\biggm|_{x=y}.
\end{equation}
This last expression implies that it is possible to make use of the Grassmann variables in the case of a non-commutative geometry.\\
Now the question will be about the form that the odd function of the Definition \ref{prod} must be. At this time we don not have an answer, but surely we can say that the form of this odd function $F(z)$ must be such as it will generate an interesting algebraic structure and of course also obey criteria that belong to the physical theories.
%
%
\section{\sc Conclusion.}
The construction of Grassmann numbers as was done in this work shows that the behavior of Definition \ref{grass} in terms of 
two binary operations over the field $\mathbb{C}$ and the $2\times2$ matrices lead us to contradictions since we were interested in
preserve both, the definition of the Grassmann variables and the equation of the Grassmannian polynomial, at the same time.\\ 
This motivates us to make a new definition, now over the binary operations of the set $\mathbb{C}$ in order to be consistent with the initial definition of the Grassmann variable. This construction only makes a difference among the Grassmann and complex numbers just in the way they make use of the binary operation, in other words they represent the same set of numbers but they do not form the same algebraic structure. It means that there is not problem considering $\mathbb{G} \times\mathbb{G}\mapsto\mathbb{C}$ (fermion-fermion), $\mathbb{C}\times\mathbb{G}\mapsto\mathbb{G}$ (boson-fermion) and $\mathbb{C}\times\mathbb{C}\mapsto\mathbb{C}$ (boson-boson) since we are working with the same set of numbers.\\
Discussions about this work with professor Lev Birbrair (Universidade Federal do Ceará) pointed us that this work could be related with the Operator Theory \cite{lev,non}.

\end{document}